\documentclass[prb,superscriptaddress,twocolumn]{revtex4-1}
\usepackage{times,amsmath}
\usepackage{epsfig}
\usepackage{color}
\usepackage{longtable}
\usepackage{graphicx}
\usepackage{dcolumn}
\usepackage{bm}
\usepackage{bookmark}
\usepackage{tabularx}
\usepackage{hyperref}
\usepackage{multirow}
\hypersetup{colorlinks=true, citecolor=blue, filecolor=blue, linkcolor=blue, urlcolor=blue}

\begin{document}

\title{Switchable Atomically Thin 2D Electrides from First-principles Prediction}

\author{Xuhui Yang}
\affiliation{College of Materials Science and Engineering, Fuzhou University, Fuzhou, Fujian Province, 350108, P. R. China}
\affiliation{Department of Physics and Astronomy, University of Nevada, Las Vegas, NV 89154,USA}

\author{Kevin Parrish}
\affiliation{Department of Physics and Astronomy, University of Nevada, Las Vegas, NV 89154,USA}

\author{Yan-Ling Li}
\affiliation{Laboratory for Quantum Design of Functional Materials, School of Physics and Electronic  Engineering, Jiangsu Normal University, Xuzhou, Jiangsu Province, 221116, P. R. China}

\author{Baisheng Sa}
\affiliation{College of Materials Science and Engineering, Fuzhou University, Fuzhou, Fujian Province, 350108, P. R. China}

\author{Hongbing Zhan}
\affiliation{College of Materials Science and Engineering, Fuzhou University, Fuzhou, Fujian Province, 350108, P. R. China}

\author{Qiang Zhu}
\email{qiang.zhu@unlv.edu}
\affiliation{Department of Physics and Astronomy, University of Nevada, Las Vegas, NV 89154,USA}

\date{\today}

\begin{abstract}
Electrides, with excess anionic electrons confined in their empty space, are promising for uses in catalysis, nonlinear optics and spin-electronics. However, the application of electrides is limited by their high chemical reactivity with the environmental agents. In this work, we report the discovery of a group of two-dimensional (2D) moonolayer electrides with the presence of switchable nearly free electron (NFE) states in their electronic structures. Unlike conventional electrides, which are metals with floating electrons forming the partially occupied bands close to the Fermi level, the switchable electrides are chemically much less active semiconductors holding the NFE states that are 0.3-1.5 eV above the Fermi level. According to a high throughput search, we identified 12 2D candidates that possess such low-energy NFE states. Among them, 11 2D materials can likely be exfoliated from the known layered materials. Under external forces, such as a compressive strain, these NFE states stemming from the surface image potential will be pushed downward to cross the Fermi level. Remarkably, the critical semiconductor-metal transition can be achieved by a strain as low as 3\% in 2D monolayer Na$_2$Pd$_3$O$_4$. As such, the switchable 2D electrides may provide an ideal platform for exploring novel quantum phenomena and modern electronic device applications.
\end{abstract}


\vskip 300 pt

\maketitle

\section{Introduction}

Electrides are a class of unconventional compounds that contain excess valence electrons confined in the void or interlayer space that play the role of anions \cite{dye2009electrides}. Chemically, such materials could serve as strong reducing agents or catalysts for chemical syntheses. From a physics point of view, the confined electrons form (partially occupied) bands close to Fermi level, which could lead to a dramatically reduced work function and high electric conductivity. These attractive physical and chemical properties collectively promise many technological applications such as the splitting of carbon dioxide at room temperature \cite{Toda-NC-2013}, synthesis of ammonia from atmospheric nitrogen under mild conditions \cite{Kitano-NChem-2012} and many others \cite{Hosono-PNAS-2017}. Recently, it was also proposed that the confined electrons in the electrides are favorable for achieving band inversions needed for topological phase transition in the electronic states \cite{Hirayama-PRX-2018,Huang-NanoL-2018, Park-PRL-2018, Zhu-PRM-2019}.

To date, a number of electride materials have been investigated both experimentally and computationally. Depending on the connectivity of crystal cavities and channels, the identified electrides could be classified to zero, one and two dimensions (0D, 1D and 2D). Of them, the 2D layered electride materials \cite{Lee-Nature-2013, Inoshita-PRX-2014}, with anionic electrons distributed in the 2D space, are of great interests since this is naturally connected with the 2D materials physics \cite{zhao2014Ca2N}. Indeed, the recent report on the synthesis of the monolayer Ca$_2$N (exfoliated from the parent layered form) has driven electride research into the nano-regime \cite{2D-Ca2N}. The monolayer Ca$_2$N, by combining high surface areas of 2D materials with the exotic properties of anionic electrons, are favorable for applications as transparent conductors in the modern electronic devices \cite{2D-Ca2N, zhao2014Ca2N}.

Despite the many unique properties that the floating anionic electrons bring, they also result in poor stabilities at room temperature and ambient atmosphere, which significantly limit the utilization of electrides as materials. In electride design, the conventional wisdom is to search for candidates with unpaired electrons present in their crystal voids \cite{Tada-IC-2014,burton2018high,Zhang-PRX-2017,ZHU20191293}. 
In the electronic band structure, these confined electrons form the distinct interstitial bands crossing the Fermi level. To date, most known electrides are strong reducing metals due to the presence of these characteristic bands. Consequently, they are unstable in water and air. From an application perspective, it is desirable to have the material remain stable under normal conditions and switch to the electride phase under the operation. Therefore, an ideal material should be semiconducting with the lowest conduction bands occupied by the interstitial electrons. Under the working condition, the interstitial band can be then downshifted to the Fermi level. As such, the interstitial band can act as an effective transport agent in electronic devices. To achieve this goal, the key is to identify the materials that can hold an interstitial band close to the Fermi level.

In many known electrides, the interstitial electrons follow the nearly free electron (NFE) model \cite{ZHU20191293} in which the electrons are considered as plane waves that are weakly perturbed by the periodic lattice potential. The dispersion can be roughly described by a parabolic-like dispersion of $E=\hbar^2k^2/2m$. While the NFE states can be found in the 3D systems, it is more natural to study them in 2D materials ~\cite{NFE-graphene,chen2012band,Cuong_2014,BL-InSe-E}. To our knowledge, most of the previous reports demonstrate that the NFE states are several eVs above the Fermi level. The high energy NFE states require common band structure engineering approaches, such as chemical/electron doping \cite{NFE-C_Nanotube, NFE-C-BN_nanotube}, strain engineering \cite{ni2012tunable, song2016room, Zhu-2013-MoS2}, and electric fields \cite{min2007ab, mak2009observation, liu2015switching, blueP-E, InSe2015JPCL, BL-InSe-E, MoS2-E, WSe2-E}, to lower their energy levels. It was only recently reported that the NFE states can be energetically found near the Fermi levels for a few 2D transition metal carbides \cite{NFE-PRB}. Based on first-principle calculations, the authors found that Sc$_2$C(OH)$_2$ was semiconducting with its conduction band occupied by the NFE states 0.45-0.94 eV above the Fermi Level \cite{NFE-JMCC}. They further proposed that the NFE states can be downshifted by external electric fields \cite{NFE-JMCC}. However, Sc$_2$C(OH)$_2$ is a hypothetical compound. To our knowledge, such switchable 2D electride phases have not been reported based on any known material.

In this work, we focus on potential 2D materials that possess accessible NFE states from strain engineering. We performed a high-throughput screening of the available 2D semiconducting materials database by searching for characteristic NFE states. From the screening, we identified 12 2D materials promising to hold the low energy NFE states at their conduction bands. Furthermore, we found that applying compressive strain can effectively push the NFE bands to the Fermi level. Different from the previous report on a hypothetical material \cite{NFE-JMCC}, 11 of the identified 2D materials can likely be exfoliated from the available layered compounds without further surface functionalization. Therefore, our predictions can be readily validated by experiment if one of the materials can be synthesized in the laboratory. Below, we will discuss the computational methods used and present our discovery in the following sections.

\begin{figure}[ht]
\centering
\includegraphics[width=0.48 \textwidth]{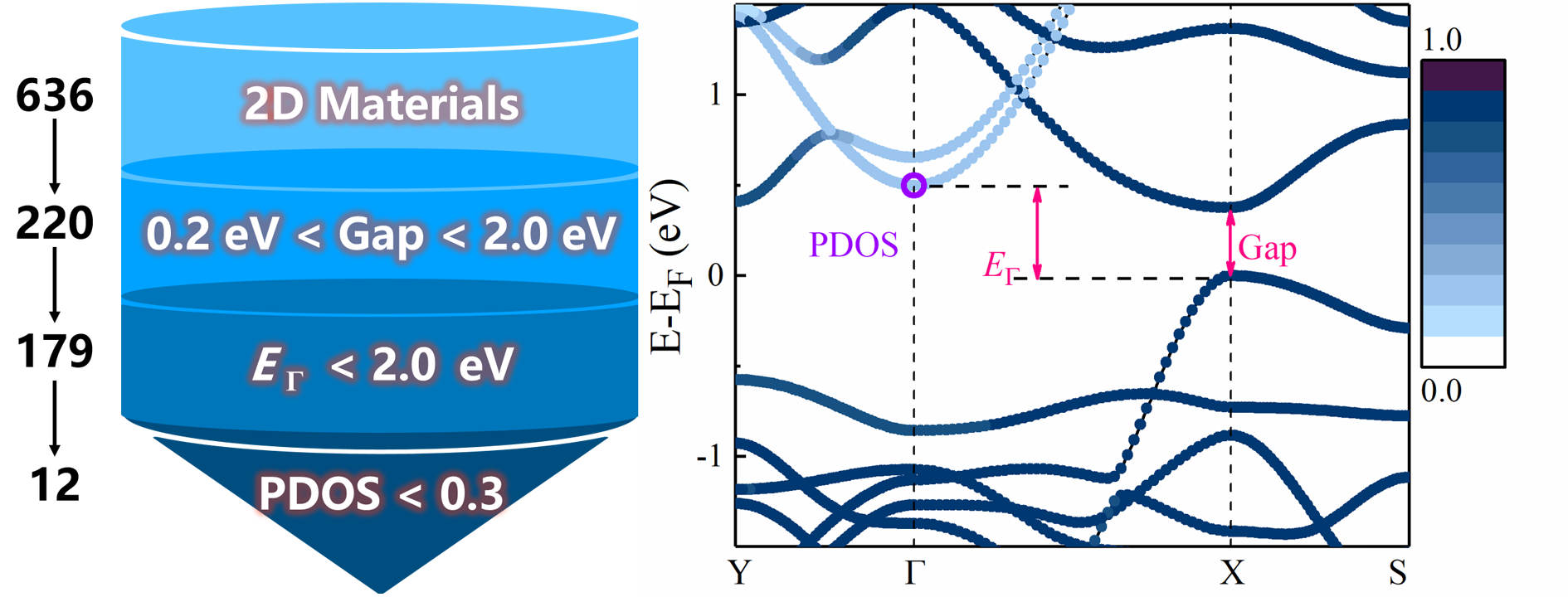}
\caption{\label{Fig1} (a) Computational Scheme for the screening of 2D swichable electrides. (b) The definition of physical quantities used in the screening. The PDOS denotes the ratio of densities which can be projected to the atomic orbitals. If the densities cannot be projected to any atomic orbitals, it suggests the electrons are loosely bound with the nucleus.}
\end{figure}

In total, 636 2D candidates from material databases were investigated based on the four steps of screening as shown in Fig.~\ref{Fig1}. The first step was to obtain potential 2D insulators/semiconductors from the online JARVIS-DFT database \cite{choudhary2017high}. For each 2D material, we performed the density functional theory (DFT) calculations using a projector augmented-wave method~\cite{PAW-PRB-1994} as implemented in the Vienna ab initio Simulation Package (VASP)~\cite{Vasp-PRB-1996}.  The exchange and correlation potentials were treated with the generalized gradient approximation (GGA) as proposed by Perdew–Burk–Ernzerhof (PBE)~\cite{PBE-PRL-1996}. Preparation of input parameters and data extraction was done by the Python Materials Genomics (Pymatgen) package~\cite{ong2013python}. We performed the geometry relaxation (MPRelaxSet), static (MPStaticSet) and non-self-consistent (MPNonSCFSet) calculations for each structure. A plane-wave cut-off energy of 700 eV was employed. The relaxation convergence criteria were 10$^{-6}$ eV for energy and 0.01 eV/\AA~ for force. In the relaxation, static and non-self-consistent calculations, the reciprocal density was set to be 300. To avoid the interlayer interaction due to periodic boundary conditions, a vacuum layer of at least 15 \AA~ thickness was added to each simulation at the screening stage. The second step was to extract the 2D semiconducting materials by applying the band gap restriction (between 0.2-2.0 eV). Steps 3 and 4 involve the identification of the NFE states. We filtered the materials with eigenstates at the $\Gamma$ point ($E_\Gamma$) of the first conduction band that are less than 2.0 eV. To ensure the conduction bands held the true NFE states, we also checked that the majority of the density of states (PDOS) could not be projected to the atomic orbitals, similar to a recent high-throughput screening work on electrides \cite{burton2018high}. After a few rounds of parameter tuning, we decided to use 0.3 as the upper bound of the PDOS value at the $\Gamma$ point as determining criteria for whether or not it is an NFE state. These screenings successfully identified 12 materials as summarized in Table \ref{all}. To check the electronic band structure further clearly, we recalculated these 12 materials with higher reciprocal density. 

\begin{table}[ht]
\caption{The summary of 2D materials investigated in this work. $E_\textrm{NFE}$ denotes the lowest gap between the NFE states and Fermi energy. $\epsilon_c$ denotes the critical biaxial compressive strain values to turn the material metallic. The exfoliation energies are extracted from the online database \cite{choudhary2017high}. Since PbS did not show zero gap even when put under up to a 20\% strain, we did not show its $\epsilon_c$ value.
}\label{all}
\begin{tabular}{ccccccc}    
\hline\hline
Material             &~~Gap~~  &$E_\textrm{NFE}$ & VBM at $\Gamma$ & ~~~~$\epsilon_c$~~~~   & $E_\textrm{exfoliation}$ \\
                     & ({eV})       & (eV)  &   & &(meV/atom)              \\
\hline
RbLiS                & 1.228   & 1.228    & Yes    & -13\%     & 94.14   \\
RbLiSe               & 1.326   & 1.326    & Yes    & -12\%     & 100.08  \\
KAgSe                & 0.511   & 0.511    & Yes    & -9\%      & 101.49  \\
K$_4$HgAs$_2$        & 0.793.  & 0.793    & Yes    & -12\%     & 107.36  \\
KMgSb                & 0.905.  & 0.905    & Yes    & -9\%      & 110.65  \\
NaZnAs               & 0.480.  & 0.480    & Yes    & -6\%      & 118.68  \\
NaZnP                & 0.719.  & 0.719    & Yes    & -7\%      & 124.80  \\
Na$_2$Pd$_3$O$_4$    & 0.376.  & 0.497    & No     & -3\%      & 139.05  \\
PbS                  & 1.654   & 1.924    & No     & N/A       & 187.54  \\
Na$_3$As             & 0.285.  & 0.285    & Yes    & -6\%      & 250.04   \\
Rb$_2$Te             & 0.477   & 0.477    & Yes    & -7\%      & 388.51  \\
SiS                  & 0.494.  & 0.494    & Yes    & -13\%     & N/A   \\
\hline\hline
\end{tabular}
\end{table}

\begin{figure}[ht]
\centering
\includegraphics[width=0.49 \textwidth]{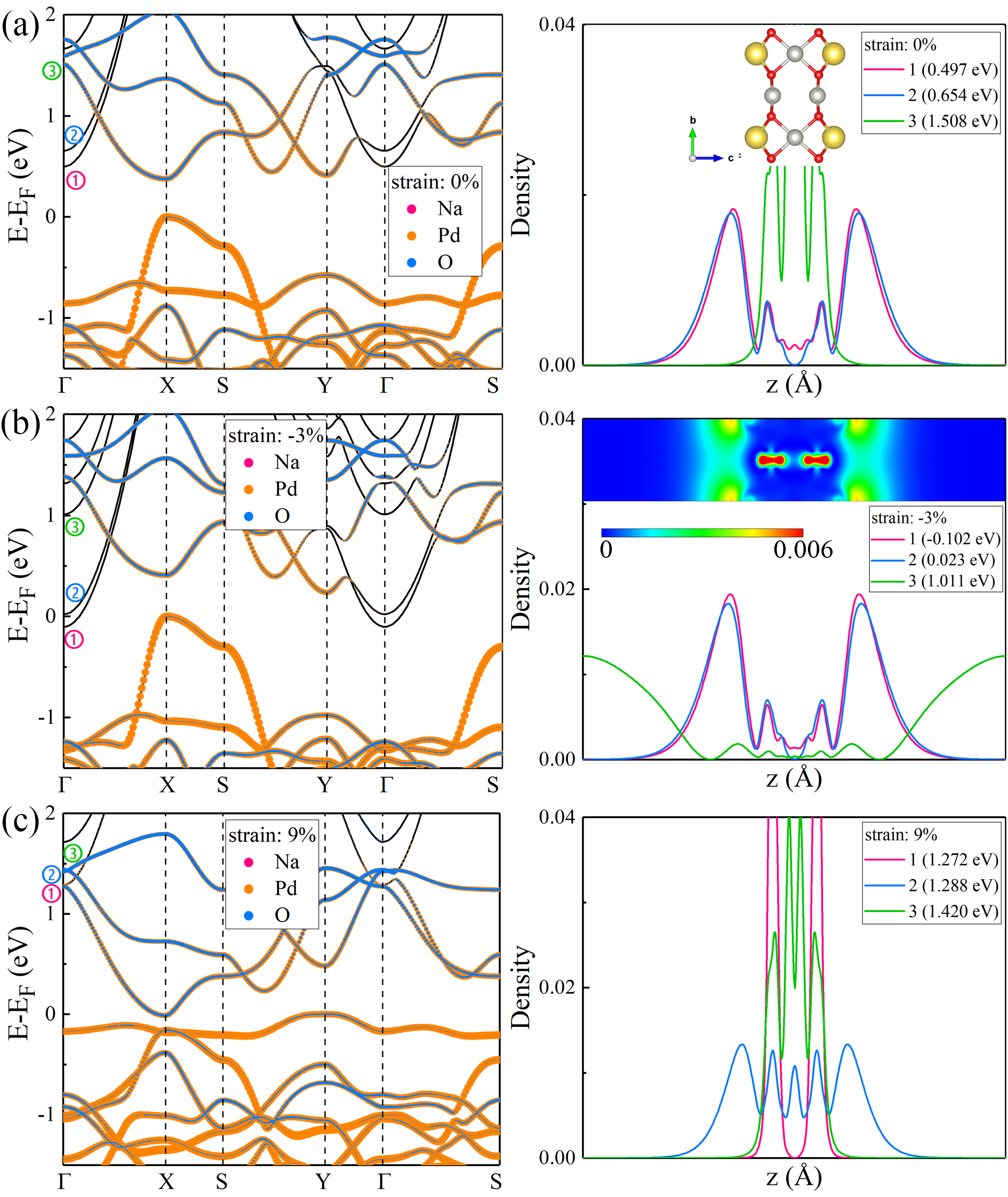}
\caption{\label{Fig2} The electronic structures of monolayer Na$_2$Pd$_3$O$_4$. (a)-(c) show the band structures and corresponding electron densities at $\Gamma$ point with 0, -3\%, 9\% strains. In the left column, each energy band is colored by the atoms which hold the largest portion of PDOS. If it cannot be projected to any atomic site, this corresponds to NFE states as denoted by black color. The right columns display the partial charge density distributions of the three lowest conduction bands at $\Gamma$ point along the $z$-axis. The order of band number are corresponding to the circled numbers in left columns. The side views of the structural model are shown in (a) as the inset. The Na, Pd and O atoms are denoted by yellow, grey and red spheres respectively. The inset in (b) is the 2D band decomposed charge density distribution of the lowest conduction band at $\Gamma$ point with the strain.}
\end{figure}

From the 12 identified materials, it is interesting to note that most of them have the alkaline metals (Li, Na, K, Rb) sitting at the outmost layer of the slab, as shown in Fig. S1 \cite{SI}. This is expected since their valence electrons are much more weakly bound than those in inner shells. To proceed, we first ruled out the materials with high exfoliation energies. Among them, PbS, Na$_3$As, Rb$_2$Te were discarded. We did not consider SiS either since this compound is from a theoretical prediction \cite{yang2016two}. Then, we applied a series of biaxial strains on the remaining compounds. Below, we will discuss how to modulate the NFE states by strain engineering on two representative materials: Na$_2$Pd$_3$O$_4$ and NaZnAs.

Fig. \ref{Fig2} displays the changes of the electronic band structures of Na$_2$Pd$_3$O$_4$ under a series of lattice strains. The pristine monolayer Na$_2$Pd$_3$O$_4$ has an orthorhombic unit cell. It is a semiconductor with a direct gap of 0.377 eV that has both the valence band maximum (VBM) and conduction band minimum (CBM) at the X point (0.5, 0, 0). Not surprisingly, the system becomes metallic when the strains are applied. However, the behaviors are drastically different under different types of strains. Under tension, the band gap goes down continuously and becomes closed when it reaches 9\% strain. On the other hand, compressive strain leads to a more dramatic change and it needs only -3\% strain to close the band gap. To understand the distinct behavior, we tracked the electronic band evolution and analyzed the characteristic wavefunctions under 0, -3\% and 9\% strains. As shown in Fig. \ref{Fig2}c, the dispersion of several conduction bands around the $\Gamma$ point follows a parabolic-like relation. We also plotted the band decomposed charge density of the three lowest conduction bands (labelled from low to high) at the $\Gamma$ point. Both bands 1 and 2 have four spikes. There are two smaller charge peaks at the oxygen's nuclear sites, suggesting that O atoms can take some electrons since they are more electronnegative than Na and Pd. More importantly, the majority of charges are symmetrically distributed outside the slab, confirming the presence of NFE states. With compression biaxial strain, the NFE states (0.497 eV at band 1 and 0.654 eV at band 2) drop more quickly than other conduction bands. At -3\% strain, the CBM is shifted to the $\Gamma$ point due to the lowered NFE states while the original CBM state at the X point remains intact. When the tensile strains are applied, the semicondutor-to-metal transition follows a completely different path in which one nearly flat valence band moves upward and the conduction bands are collectively pushed downward. Due to these joint efforts, the gap is finally bridged. However, it is notable that the NFE states actually take higher energies during the metallization process. The total evolution has been shown in Fig. S3 \cite{SI}.

\begin{figure}
\centering
\includegraphics[width=0.48 \textwidth]{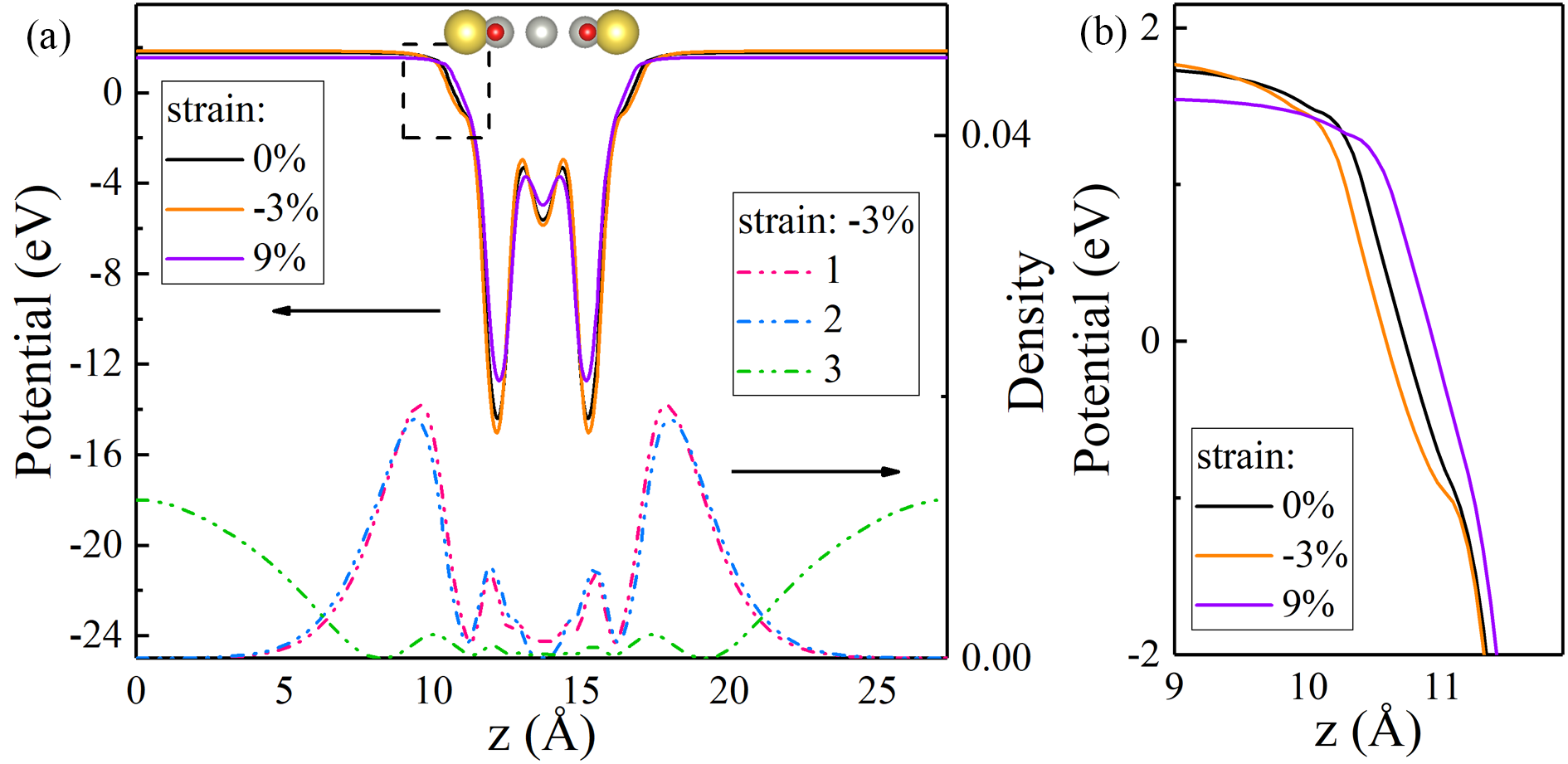}
\caption{\label{Fig3} (a) The calculated electrostatic potential distribution of monolayer Na$_2$Pd$_3$O$_4$ with/without biaxial strains. The band decomposed charge density distribution at $\Gamma$ point of the first conduction band under -3 \% biaxial strain is also shown for comparison. The positions of each atom along $z$-axis are shown as inset. The Na, Pd and O atoms are denoted by yellow, grey and red spheres respectively. (b) The enlarged detail of the dashed box in (a).}
\end{figure}

\begin{figure}
\centering
\includegraphics[width=0.48 \textwidth]{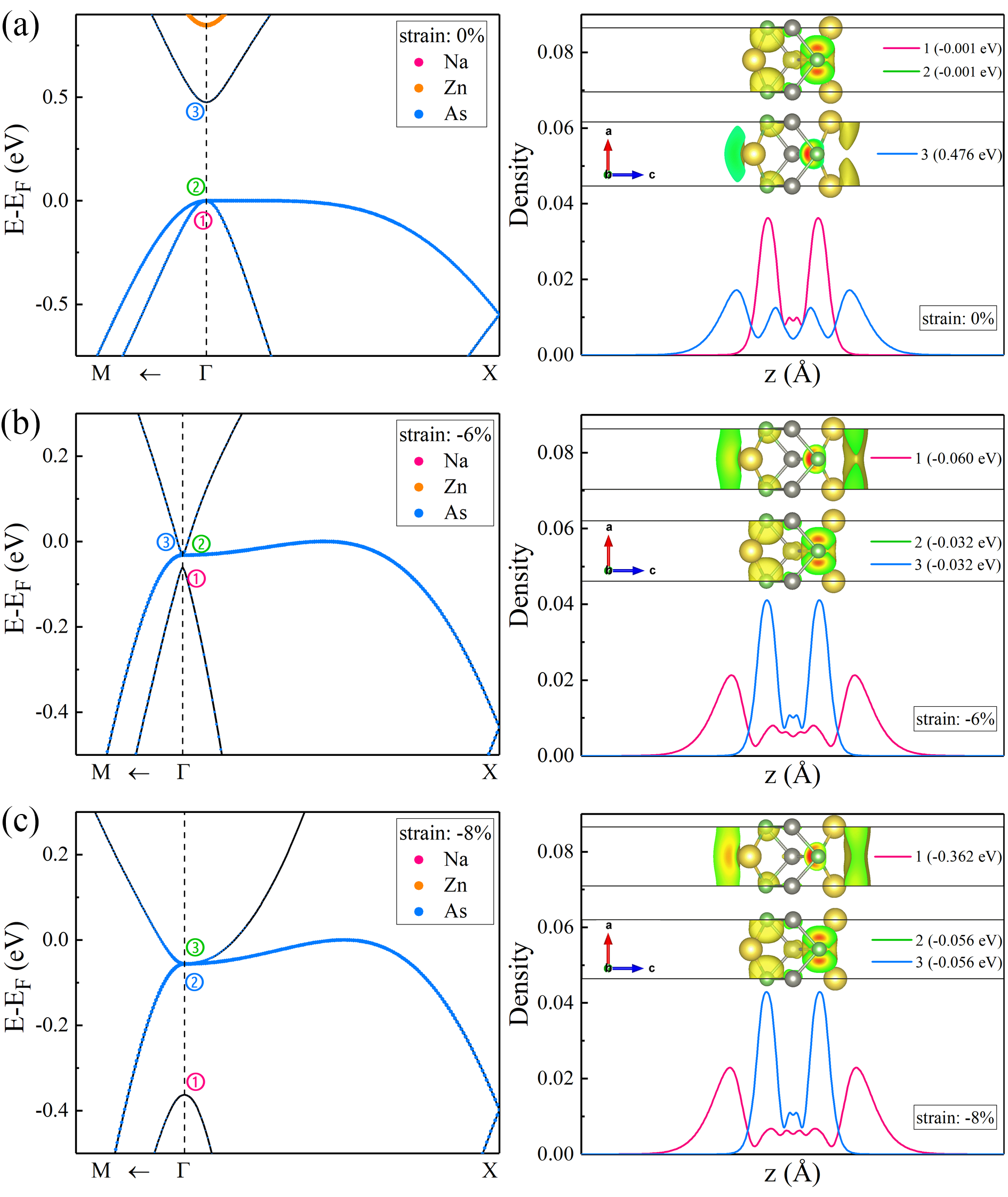}
\caption{\label{Fig4} The electronic structures of monolayer NaZnAs. The band structures under (a) 0, (b) -6\% and (c) -8\% biaxial strains. The right column plots the corresponding partial charge densities of three representative bands (two highest valence bands and one lowest conduction band) at $\Gamma$ point along the $z$-axis. For comparison, the isosurfaces of charge density plots for each band were also shown as the insets. The Na, Zn and As atoms are denoted by yellow, grey and green spheres respectively.}
\end{figure}

The direction dependent behavior can be understood by examining the nature of the NFE states. We first found that the bulk Na$_2$Pd$_3$O$_4$ did not exhibit low-energy NFE bands (Fig. S2) \cite{SI}. Different from the bulk electrides where the NFE states are occupying the crystal voids, the NFE states of 2D electrides stem from the image potential states. When the metal has a band gap near the vacuum level $E_\textrm{vacuum}$, an electron below $E_\textrm{vacuum}$ may be trapped in the potential well formed by the Coulomb-like attractive image potential and the repulsive surface barrier \cite{hofer1997time, NFE-C_Nanotube, NFE-C-BN_nanotube, NFE-graphene}. Whether or not the electron can be trapped depends on the spatial distribution of the image potential. If the image potential has a long tail, the NFE states are more likely to be stabilized. Fig. \ref{Fig3} displays the electrostatic potential (ESP) distribution of Na$_2$Pd$_3$O$_4$ under 0, -3\% and 9\% strains. Clearly, the tensile strain sharpens the ESP distribution compared to the pristine form. Thus, the NFE states under tension occupy higher energy levels compared to those in the pristine form. On the other hand, the compressive strain makes the ESP more extended by creating a shoulder around -1 eV. Consequently, the electrons can be trapped around the shoulder with a lower eigenvalue. We also confirmed the results by numerical simulation. Solving the 1D Schordinger equation for the potential wells given by Fig. \ref{Fig3}, we indeed found that the NFE states under the negative strain yielded lower eigenvalues as compared to those under the zero strain (Fig. S4-S5) \cite{SI}. Therefore, we conclude that a negative biaxial strain can promote the activity of the NFE states by modifying the ESP distribution. 

Different from Na$_2$Pd$_3$O$_4$, many other materials in Table \ref{all} have their VBM at the $\Gamma$ point. For these materials, the semiconductor-metal transitions under strain become more complicated. As shown in Fig. \ref{Fig4}, the pristine monolayer NaZnAs has its VBM at the $\Gamma$ point. This VBM corresponds to the As's $p_x$/$p_y$ orbitals (bands 1 and 2, see more details in Fig. S6 \cite{SI}). The CBM at the $\Gamma$ point of band 3 is 0.476 eV above the VBM. The decomposed charge density plot suggests that four competing maxima exist in band 3 with two corresponding to the NFE states and the other two corresponding to the electrons trapped by the As nucleus. Compared to band 1 of the Na$_2$Pd$_3$O$_4$ in Fig. \ref{Fig2}a, the dispersion of band 3 in Fig. \ref{Fig4}a is less parabolic since it has more densities inside the slab. Clearly, the hybridization between As's $p_x$/$p_y$ orbitals (bands 1 and 2) and the interstitial electrons dominate the electronic properties near the Fermi level. When put under compressive strain, the energy of NFE states got closer to the Fermi level. Consequently, there is a prominent charge transfer from As's $p_x$/$p_y$ orbitals to the interstitial electrons. When it reaches a critical strain near -6\% (Fig. \ref{Fig4}b), the charge transfer is complete and the hybridization effect becomes negligible. Therefore, further applying the strain leads to a quick downshift of the interstitial electrons (band 1 in Fig. \ref{Fig4}c). For instance, at -8\% strain, the energy of interstitial electrons was reduced to -0.362 eV, while the Fermi level is occupied by As's $p_x$/$p_y$ orbitals (bands 2 and 3).
These phenomena suggest that these 2D materials with low-energy NFE states can exhibit very rich electronic behavior by altering their mechanical attributes. In particular, the coexistence of flat band and NFE band (Fig. \ref{Fig4}b) may be useful for exploring novel quantum phenomena.

\begin{figure}
\centering
\includegraphics[width=0.35 \textwidth]{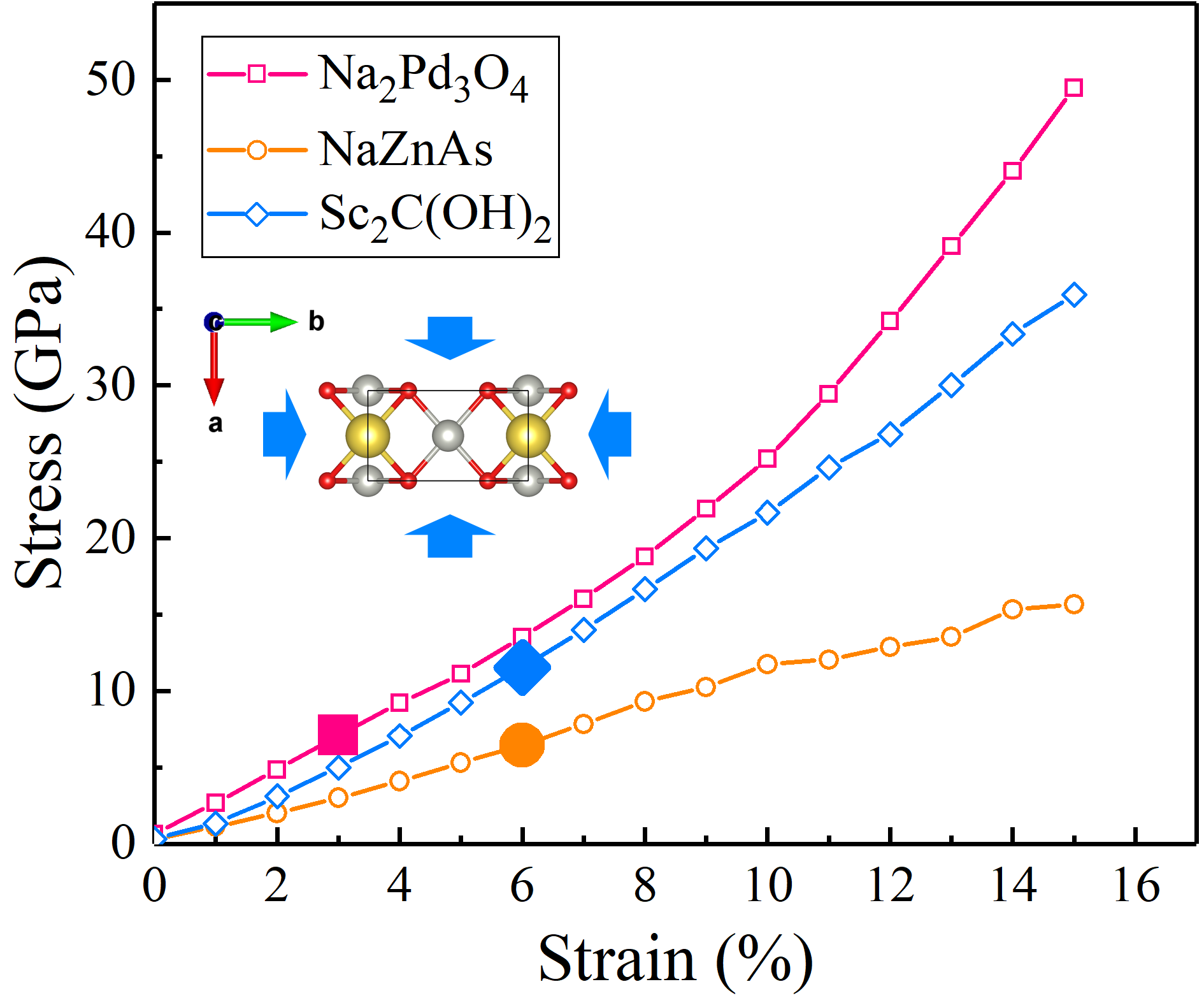}
\caption{\label{Fig5} Calculated stress–strain curves of monolayer Na$_2$Pd$_3$O$_4$. The inset is the schematic diagram of biaxial compressive strain. The solid points denote the critical biaxial compressive strain values $\epsilon_c$ to complete the semiconductor-metal transition.}
\end{figure}

We also checked the effects of strain on the previously reported Sc$_2$C(OH)$_2$ (Fig. S7) \cite{SI}. It showed a similar semiconductor-metal transition at the -6\% biaxial strain, suggesting that it's easier to trigger the electride phase transition in Na$_2$Pd$_3$O$_4$ than in Sc$_2$C(OH)$_2$. From Fig. \ref{Fig5}, it is clear that the stress-strain relations follow a linear behavior up to the -8\% strain. The -3\% compressive strain falls into the region of elastic deformation. Therefore, the transition between the electride and non-electride phase of Na$_2$Pd$_3$O$_4$ is likely reversible, which is advantageous for switchable device applications. We also systematically investigated other materials in Table \ref{all}. Among them, 5 materials need a critical strain less than 10\% to complete the semiconducting to metal transition. Introducing strains to 2D nanomaterials has been widely applied to tune the material’s properties \cite{deng2018strain}. These materials require energies between 90-130 meV/atom to be exfoliated from their parent materials, which is higher than graphene (70 meV/atom) and MoS$_2$ (76 meV/atom) but comparable to PtSe$_2$ (111 meV/atom) \cite{wang2015monolayer, choudhary2017high}. To our knowledge, none of the materials in Table \ref{all} have been experimentally synthesized. However, several of them have been theoretically investigated \cite{yang2016two, wang2018kagse, zhang2020high}. If one of the 2D forms can be made, their electric measurements should be straightforward. 

In sum, we report the computational design of 2D switchable atomically thin electrides from first-principles calculations in conjunction with high-throughput screening. We have investigated over 600 potential 2D materials which are potentially exfoliatable from the existing layered materials. Among them, we found that a family of 2D materials may hold the NFE states that are close to the Fermi level. Different from the conventional bulk electrides or the recently synthesized 2D monolayer intrincic electride Ca$_2$N, these monolayer electrides are more chemically inert semiconductors at normal conditions. Under a strain less than 10\%, their NFE states can be lowered and form the partially occupied bands crossing the Fermi level. Consequently, the electride phase transition can be reversible via strain engineering, which provides an ideal platform for switchable nano device/sensors for gas detection and electron transports. Compared to recent studies \cite{NFE-JMCC}, the parent materials discussed in this work have already been synthesized, and several of them require less critical strain to activate the electride phase. In addition to strain manipulation, other band structure engineering methods, such as applying the electric field and chemical doping, are also expected to work given that most of these materials have NFE states close to the Fermi level. Thanks to the improved stability, these predicted materials are more amenable for materials fabrication and operation. We hope this work can stimulate further experimental investigation.

QZ and XHY acknowledge the use of computing resources from XSEDE (DMR180040). This work was also supported by the National Natural Science Foundation of China (Grant Nos. 51872048, 21973012). The study of XHY at University of Nevada is supported by China Scholarship Council (CSC) No.201806650022.

\bibliography{ref}

\end{document}